\documentclass[aps,prb,amsmath,amssymb,floatfix,twocolumn]{revtex4-1}
\usepackage{bbold}
\usepackage{graphicx}
\usepackage{color}
\usepackage{hyperref}
\usepackage{wasysym}

\begin{document}

\title{Charge-$2e$ Skyrmion condensate in a hidden order state }
\author{Chen-Hsuan Hsu}
\affiliation{Department of Physics and Astronomy, University of
California Los Angeles\\ Los Angeles, California 90095-1547}
\author{Sudip Chakravarty}
\affiliation{Department of Physics and Astronomy, University of
California Los Angeles\\ Los Angeles, California 90095-1547}

\date{\today}

\begin{abstract}
A higher angular momentum ($\ell =2$) $d$-density wave, a mixed triplet and a singlet, interestingly, admits skyrmionic textures. The Skyrmions carry charge $2e$ and can condense into a spin-singlet $s$-wave superconducting state. In addition, a charge current can be induced by a time-dependent inhomogeneous spin texture, leading to quantized charge pumping. The quantum phase transition between this mixed triplet $d$-density wave and skyrmionic superconducting condensate likely leads to  deconfined quantum critical points. We suggest connections of this exotic state to electronic materials that are strongly correlated, such as the heavy fermion $\mathrm{URu_{2}Si_{2}}$. At the very least, we provide a concrete example in which topological order and broken symmetry are intertwined, which can give rise to  non-BCS superconductivity.
\end{abstract}


\maketitle

\section{Introduction}
It has become very much in vogue to argue that topological aspects of condensed matter bear no relation to  broken symmetries.~\cite{Qi:2011,*Hasan:2010} In a strict sense 
this need not be so.~\cite{Raghu:2008,*Sun:2009,*Yang:2010} One can construct examples where a broken symmetry state has interesting topological properties and   can even be
protected by the broken symmetry itself. An interesting  example of a mixed triplet $d$-density wave and its possible relevance to one of the many competing phases in the high temperature cuprate phase digram was recently demonstrated,~\cite{Hsu:2011} where   it was found that the system exhibits quantized spin Hall effect
even without any explicit  spin-orbit coupling. In the present paper we show that the system exhibits charge $2e$ skyrmions, which can condense into a remarkable superconducting state. As we shall discuss, such a mixed triplet $d$-density wave  system and the resulting superconductivity  is potentially relevant 
to  the heavy fermion $\mathrm{URu_{2}Si_{2}}$ with hidden order.~\cite{Mydosh:2011}

An early attempt at such a non-BCS mechanism   of superconductivity  was made  by Wiegmann,~\cite{Wiegmann:1999} as an extension of Fr\"ohlich mechanism to higher dimension. More recently,  several interesting papers have  led to discussions of superconductivity  in single and bilayer graphenes. Grover and Senthil~\cite{Grover:2008} have  provided a mechanism in which  electrons hopping on a honeycomb lattice can lead to a charge-$2e$ skyrmionic condensate, possibly relevant to  single layer graphene. To a  certain degree we follow their formalism; see also the earlier  work in Ref.~\onlinecite{Sondhi:1993} of charge-$e$ skyrmions in a quantum Hall ferromagnet. As to  bilayer graphene,  a charge-$4e$ skyrmionic condensate  has been suggested by Lu and Herbut~\cite{Lu:2012} and Moon.~\cite{Moon:2012} In the present paper we base our results, instead,  on an unusual {\em spontaneously broken symmetry}  generated by electron-electron interaction, not  by a given non-interacting band structure of a  material, namely the mixed triplet-singlet density wave state of angular momentum $\ell = 2$,  and we point out its possible implications to the mysterious hidden order state in   $\mathrm{URu_{2}Si_{2}}$, in particular to its superconductivity.

It is appropriate to comment on what we mean by ``hidden order''. An order parameter can often be inferred from its macroscopic consequences in terms of  certain generalized rigidities. Sometimes its direct microscopic signature
is difficult to detect: a direct determination of superconducting order, a broken global $U(1)$ gauge symmetry, requires subtle Josephson effect, and even antiferromagnetic order requires microscopic neutron scattering probes.
Density wave states of higher angular momentum, such as the mixed triplet $d$-density wave,  are even harder to detect. It does not lead to a net charge density wave  or spin density wave to be detected by  common $s$-wave probes. It is further undetectable because it does not even break time-reversal invariance. A discussion of possible experimental detections of particle-hole condensates of higher angular momentum was given in Ref.~\onlinecite{Nayak:2000}. Thus, it is fair to conclude that the state we consider here is a good candidate for a hidden order.

It is also necessary to remark on the realization of particle-hole condensates of higher angular momentum. An effective low energy theory of a strongly correlated system is bound to have a multitude of coupling
constants, perhaps hierarchically arranged. In such cases, we can generally expect a phase diagram with a multitude of broken symmetry states. It is a profound mystery as to why non-trivial examples are so few and far between. A partial reason could be, as stated above,   that these states are unresponsive to  common
$s$-wave probes employed in condensed matter physics and therefore appear to be hidden.~\cite{Chakravarty:2001}

The next question is: are these low energy effective Hamiltonians contrived? If so, it would be of value to pursue them. However, simple Hartree-Fock analyses have shown that they certainly are not:~\cite{Nayak:2000,Nersesyan:1991} an onsite repulsion $U$, a nearest neighbor interaction $V$, and an exchange interaction $J$ are sufficient in a single band model. 

The structure of this paper is as follows: in Sec. \ref{Sec:action}, we construct the low energy effective action of the mixed triplet and singlet $d$-density wave system. In Sec. \ref{Sec:charge}, we compute the charge and the spin of a skyrmion and verify that the skyrmions in this system are bosons, which can lead to a superconducting phase transition. In Sec. \ref{Sec:angular}, we compute the angular momentum of a skyrmion. In Sec. \ref{Sec:pumping}, we study the charge pumping due to a time-dependent inhomogeneous spin texture that is interesting in its own right. In Sec.~\ref{discussion} we  discuss mainly the problem of $\mathrm{URu_{2}Si_{2}}$. In the appendices,  the derivation of the non-linear $\sigma$-model and the details of computing the Chern-Simons coefficients and charge pumping are provided.

\section{\label{Sec:action}Effective action}

In the momentum space the mixed triplet and singlet $d$-density wave order parameter is ($c$ and $c^{\dagger}$ are fermonic annihilation and creation operators, respectively,  $Q=(\pi,\pi)$, and the lattice constant is set to unity) 
\begin{equation}
\langle c_{k+Q,\alpha}^{\dagger} c_{k,\beta}\rangle \propto i(\vec{\sigma} \cdot \hat{N})_{\alpha\beta} W_k + \delta_{\alpha\beta} \Delta_k,
\end{equation}
where $\hat{N}$ is a unit vector, $\vec{\sigma}$ are the Pauli matrices acting on spin indices, and the form factors
\begin{eqnarray}
W_k &\equiv& \frac{W_0}{2} (\cos \it{k_x} - \cos \it{k_y}),\\
\Delta_k &\equiv& \Delta_0  \sin k_x \sin k_y,
\end{eqnarray}
correspond to the $d_{x^2-y^2}$ and $d_{xy}$ density wave, respectively~\cite{Nayak:2000}. It is not necessary that $d_{xy}$ and $d_{x^{2}-y^{2}}$ transitions be close to each  other, nor are they required to be close in energy~\cite{Hsu:2011}.

If we choose the spin quantization axis to be $\hat z$, the up spins represent circulating spin currents corresponding to the 
order parameter $d+id$ and the down spins to $d-id$ (in an abbreviated notation). So, there are net circulating spin currents alternating from one plaquette to the next but no circulating charge currents. By the choice of the quantization axis we have explicitly broken $SU(2)$, but not $U(1)$, and the coset space of the order parameter $S^{2}\equiv SU(2)/U(1)$. Such a state can admit Skyrmions in two dimensions (ignoring the possibility of hedgehog configurations in $(2+1)$ dimensions (cf. below).

The Hamiltonian is
\begin{equation}
\mathcal{H}= \sum_{k,\alpha,\beta} \psi_{k,\alpha}^{\dagger} \left[  \delta_{\alpha\beta} (\tau^{z} \epsilon_k +
 \tau^{x} \Delta_k)
- (\vec{\sigma} \cdot \hat{N})_{\alpha\beta} \tau^{y} W_k \right]\psi_{k\beta},
\end{equation}
where the summation is over the reduced Brillouin Zone (RBZ) bounded by $k_y \pm k_x = \pm \pi$, the spinor is $\psi_{k,\alpha}^{\dagger}\equiv (c_{k,\alpha}^{\dagger}, c_{k+Q,\alpha}^{\dagger})$, and $ \epsilon_k \equiv -2t (\cos k_x + \cos k_y)$; addition of longer ranged hopping will not change our conclusions~\cite{Hsu:2011}. Here $\tau^{i}$ ($i=x,y,z$) are Pauli matrices acting on the two-component spinor. It is not necessary 
but convenient to construct a low energy effective field theory. For this we expand around the points $K_1 \equiv(\frac{\pi}{2},\frac{\pi}{2})$ and $K_2 \equiv(-\frac{\pi}{2},\frac{\pi}{2})$, what would have been the two distinct nodal points in the absence of the $d_{xy}$ term, and $K_3 \equiv(0,\pi)$, what would have been the nodal point in the absence of the $d_{x^2-y^2}$ term. 
This allows us to develop an effective \text{low} energy theory by separating the fast modes from the s\text{low} modes. After that we make a sequence of transformations for simplicity: (1) transform the Hamiltonian to the real space, which al\text{low}s us to formulate the skyrmion problem; (2) perform a $\pi/2$ rotation along the $\tau^y$-direction, which al\text{low}s us to match to the notation of Ref.~\onlinecite{Nersesyan:1991} for the convenience of the reader; (3) label  $\psi_{K_i+q,\alpha}$ by $\psi_{i\alpha}$, since $K_{i}$ is now a redundant notation; (4) construct the imaginary time effective action, with the definition $\bar{\psi} \equiv -i \psi^{\dagger} \tau ^z$. Finally, after suppressing the spin indices, and with the definitions
$\gamma^0 \equiv \tau^z$, $\gamma^x \equiv \tau^y$, and $\gamma^y \equiv -\tau^x$, we obtain the effective action in a more compact notation: 
\begin{widetext}
\begin{eqnarray}
S &=& \sum_{j=1,2}\int d^3 x\; \bar{\psi}_{j} \left[ 
-i \gamma^0 \partial_{\tau}
- 2it  \gamma^x (\eta_{j}\partial_{x}+\partial_{y}) 
+ i\frac{W_0}{2} (\vec{\sigma} \cdot \hat{N}) \gamma^y 
(-\eta_{j}\partial_{x}+\partial_{y})  + i \eta_{j}\Delta_0   \right] \psi_{j} \nonumber \\
&& +\int d^3 x \bar{\psi}_{3} \left[ 
-i \gamma^0 \partial_{\tau}
- W_0 (\vec{\sigma} \cdot \hat{N}) \gamma^y \right] \psi_{3},
\end{eqnarray}
\end{widetext}
where $\eta_{1}=1$ and $\eta_{2}=-1$.

\section{\label{Sec:charge}The charge and spin of a Skyrmion}

We will compute the charge of the skyrmions in the system by following Grover and Senthil's adiabatic argument.~\cite{Grover:2008} First, consider the action around $K_1=(\frac{\pi}{2},\frac{\pi}{2})$ when the order parameter is uniform (say, $\hat{N}=\hat{z}$). The results for $K_2=(-\frac{\pi}{2},\frac{\pi}{2})$  and $K_3=(0,\pi)$
 follow identically. In our previous paper we have shown that in this case the non-trivial topology leads to a quantized spin Hall conductance in  $i\sigma d_{x^2-y^2}+d_{xy}$-density wave state~\cite{Hsu:2011} as long as the system is fully gapped. The spin quantum Hall effect implies that the external gauge fields $A^c$ and $A^s$ couple to charge and spin currents, respectively. In the presence of these external gauge fields, we add minimal coupling in the action by 
\begin{equation}
\frac{1}{i} \partial_{\mu} = p_{\mu} \rightarrow  p_{\mu} +A^c_{\mu}+\frac{\sigma^z}{2} A^s_{\mu}.
\end{equation}
Then the action is
\begin{widetext}
\begin{eqnarray}
S_1[A^c,A^s] &=& \int d^3 x \; \bar{\psi}_{1} \left[ 
-i \gamma^0 \partial_{\tau} + \gamma^0 (A^c_{\tau}+\frac{\sigma^z}{2}A^s_{\tau})
- 2it  \gamma^x (\partial_{x}+\partial_{y}) + 2t \gamma^x (A^c_{x}+\frac{\sigma^z}{2}A^s_{x} + A^c_{y}+\frac{\sigma^z}{2}A^s_{y}) \right.\nonumber 
\\
&& \left. + i\frac{W_0}{2}  \sigma^z \gamma^y (-\partial_{x}+\partial_{y})  
- \frac{W_0}{2}  \sigma^z \gamma^y (-A^c_{x}-\frac{\sigma^z}{2}A^s_{x} + A^c_{y}+\frac{\sigma^z}{2}A^s_{y})
+ i \Delta_0   \right] \psi_{1}, 
\label{Eq:min}
\end{eqnarray}
where we set $e=\hbar=1$. The non-vanishing transverse spin conductance implies that the low energy effective action for the gauge fields is given by
\begin{eqnarray}
S_{\text{eff}}=\frac{i}{2\pi} \int d^3\it{x} \epsilon^{\mu\nu\lambda}A^c_{\mu}\partial_{\nu}A^s_{\lambda},
\end{eqnarray}
and the charge current is induced by the spin gauge field
\begin{eqnarray}
j^c_{\mu}= \frac{1}{2\pi} \epsilon^{\mu\nu\lambda}\partial_{\nu}A^s_{\lambda}.
\end{eqnarray}
\end{widetext}

Consider now a static configuration of the $\hat{N}$ field with unit Pontryagin index in the polar coordinate $(r,\theta)$: 
\begin{equation}
\hat{N}(r,\theta)=\left[ \sin \alpha(r) \cos \theta, \sin \alpha(r) \sin \theta, \cos \alpha(r) \right]
\end{equation}
with the boundary conditions $\alpha(r=0)=0$ and $\alpha(r\rightarrow \infty)=\pi$. 
Performing  a unitary transformation at all points in space such that
$U^{\dagger} (\vec{\sigma} \cdot \hat{N}) U = \sigma^{z}$,
and defining $\psi=U \psi'$, and $\bar{\psi}=\bar{\psi}'U^{\dagger}$, we  obtain
\begin{widetext}
\begin{eqnarray}
S_1 
&=& \int d^3x \;  \bar{\psi}_{1}' \left[
-i \gamma^0 \partial_{\tau} - 2it  \gamma^x (\partial_{x}+\partial_{y}) 
+ i\frac{W_0}{2} \sigma^z \gamma^y (-\partial_{x}+\partial_{y})  
+ i \Delta_0 \right] \psi'_{1} \nonumber\\
&+& \int d^3 x \;  \bar{\psi}_{1}' \left[ -i \gamma^{0}( U^{\dagger}\partial_{\tau} U) -2it \gamma^x (U^{\dagger}\partial_{x}U+U^{\dagger}\partial_{y}U)  +i \frac{W_0}{2} \sigma^z \gamma^y (-U^{\dagger}\partial_{x}U+U^{\dagger}\partial_{y}U) \right] \psi'_{1}
\label{Eq:rot}
\end{eqnarray}
To proceed, we write down the explicit form for $U(r,\theta)$, which is
\begin{equation}
U(r,\theta)=\left(
\begin{array}{cc}
\cos \frac{\alpha(r)}{2} & -\sin \frac{\alpha(r)}{2} e^{-i\theta} \\
\sin \frac{\alpha(r)}{2} e^{i\theta} & \cos \frac{\alpha(r)}{2}
\end{array}
\right),
\end{equation}
In the far field limit,
$
U^{\dagger} \partial_{x} U 
=  (\frac{-i\sin \theta}{r})\sigma^z$, and 
$U^{\dagger}\partial_{y} U
=  (\frac{i\cos \theta}{r})\sigma^z 
$; 
substituting  into Eq.~(\ref{Eq:rot}) and introducing  $f_{\mu} = -i U^{\dagger} \partial_{\mu} U$, we get
\begin{eqnarray}
S_1
&=& \int d^3 x\; \bar{\psi}_{1}' \left[
-i \gamma^0 \partial_{\tau} - 2it  \gamma^x (\partial_{x}+\partial_{y}) 
+ i\frac{W_0}{2} \sigma^z \gamma^y (-\partial_{x}+\partial_{y})  
+ i \Delta_0 \right] \psi'_{1} \nonumber\\
&+& \int d^3 x \; \bar{\psi}_{1}' \left[ 2t \gamma^x 
( f_x + f_y  )  + \frac{W_0}{2} \sigma^z \gamma^y ( f_x - f_y ) \right] \psi'_{1}
\end{eqnarray}
\end{widetext}
Equating the above equation and Eq.~(\ref{Eq:min}), 
we obtain  in the far field limit
\begin{eqnarray}
A^c_{x}= A^c_{y}=0 ; \;  A^s_{x}= - \frac{2\sin \theta}{r};  A^s_{y}=\frac{2\cos \theta}{r}. \nonumber
\end{eqnarray}
In other words, the process of tuning the order parameter from $\sigma^z$ to $\hat{\sigma} \cdot \hat{N}(r,\theta)$  is equivalent  to adding an external spin gauge field
\begin{equation}
\vec{A}^s = -\frac{2\sin \theta}{r} \hat{x}+\frac{2\cos \theta}{r} \hat{y} = \frac{2}{r} \hat{\theta}.
\end{equation}
The total flux of this gauge field is clearly $4\pi$.
Suppose we adiabatically construct the Skyrmion configuration $\hat{N}(r,\theta)$ from the ground state $\hat{z}$ in a very long time period $\tau_{p}\rightarrow \infty$. During the process, we effectively thread a spin gauge flux of $4\pi$. The transverse spin Hall conductance implies that a radial current $j^c_{r}$ will be induced by the $4\pi$ spin gauge flux of $\vec{A}^s(t)$, which is now time-dependent: $\vec{A}^s(t=0)=0$ and $\vec{A}^s(t=\tau_{p})=\vec{A}^s$, that is,
\begin{equation}
j^c_{r}(t) = -\frac{1}{2\pi} \partial_t A^s_{\theta}(t).
\end{equation}
As a result, charge will be transferred from the center to the boundary, and the total charge transferred is 
\begin{equation}
Q^c=\int^{\tau_{p}}_{0} dt \int^{2\pi}_{0} r d\theta j^c_r(t)  = -2.
\end{equation}
Therefore, after restoring the unit of charge to $e$, we obtain a Skyrmion with charge $2e$; its spin is 0.  

It is important to verify the adiabatic result by a different method. This can be done by a  computation of the Chern number.~\cite{Yakovenko:1997} 
The charge and spin of the skyrmions are associated with the coefficients of the Chern-Simons terms by the following relations: $Q_{\text{skyrmion}}= C_{2}e$ and $S_{\text{skyrmion}}=C_{1} \frac{\hbar}{2}$,
where $C_{1}$ and $C_{2}$ are \begin{eqnarray}
C_{1}&=&\frac{\epsilon_{\mu\nu\lambda}}{24\pi^2} {\text{Tr}} \left[ \int d^3k 
G \frac{\partial G^{-1}}{\partial k_{\mu}} 
G \frac{\partial G^{-1}}{\partial k_{\nu}}
G \frac{\partial G^{-1}}{\partial k_{\lambda}} \right], \\
C_{2}&=&\frac{\epsilon_{\mu\nu\lambda}}{24\pi^2}  {\text{Tr}} \left[ \int d^3k (\vec{\sigma} \cdot \hat{z})
G \frac{\partial G^{-1}}{\partial k_{\mu}} 
G \frac{\partial G^{-1}}{\partial k_{\nu}}
G \frac{\partial G^{-1}}{\partial k_{\lambda}} \right],\nonumber\\
\end{eqnarray}
where $G$ is the matrix Geen's function and the trace is taken over the spin index $\sigma$ and other discrete indices.

If the Green's function matrix is diagonal in the spin index, then the Chern-Simons coefficients for up and down spins can be computed separately. 
\begin{equation}
{\cal N}(G_{\sigma})=\frac{\epsilon_{\mu\nu\lambda}}{24\pi^2} \text{Tr} \left[ \int d^3k 
G_{\sigma} \frac{\partial G_{\sigma}^{-1}}{\partial k_{\mu}} 
G_{\sigma} \frac{\partial G_{\sigma}^{-1}}{\partial k_{\nu}}
G_{\sigma} \frac{\partial G_{\sigma}^{-1}}{\partial k_{\lambda}} \right],
\end{equation}
and
$C_{1}= {\cal N}(G_{\uparrow}) + {\cal N}(G_{\downarrow})$,
$C_{2}= {\cal N}(G_{\uparrow}) - {\cal N}(G_{\downarrow})$.
Furthermore, it can be shown (see Appendix \ref{Sec:Chern}) that for
\begin{eqnarray}
 G_{\sigma}^{-1}= i \omega \hat{I} - \hat{\tau} \cdot \vec{h}_{\sigma}
\end{eqnarray}
with $\vec{h}_{\sigma}$ being the Anderson's pseudospin vector~\cite{Anderson:1958} of the Hamiltonian, the Chern-Simons coefficient for spin $\sigma$ can be written as
\begin{eqnarray}
{\cal N}(G_{\sigma})= -\int \frac{d^2k}{4\pi} 
\hat{h}_{\sigma} \cdot
\frac{\partial\hat{h}_{\sigma}}{\partial k_{x}} \times 
\frac{\partial\hat{h}_{\sigma}}{\partial k_{y}},  
\end{eqnarray}
where $\hat{h}_{\sigma} \equiv \vec{h}_{\sigma}/|\vec{h}_{\sigma}|$ is the unit vector of $\vec{h}_{\sigma}$.  Here  $C_1$ and $C_2$ are the total Chern number and the spin Chern number $\cal{ N}_{\text{spin}}$ defined in our previous paper, respectively.~\cite{Hsu:2011} For $i\sigma d_{x^2-y^2}+d_{xy}$ system, we have $\vec{h}_{\sigma} \equiv (\Delta_k, -\sigma W_k,\epsilon_k)$. Explicitly, $C_{1}= -1 + 1 = 0$ and $C_{2}= -1 - 1 = -2$;  thus the results are the same as above.

Because a Skyrmion in the system carries integer spin, it obeys bosonic statistics and may undergo Bose-Einstein condensate. As a result, the charge-$2e$ Skyrmion condensate will lead to a superconducting phase transition.  But what about its orbital angular momentum? In the following  section, we will prove that it is zero resulting in a $s$-wave singlet state. This is a bit surprising given the original $d$-wave form factor. 

\section{\label{Sec:angular}The angular momentum of a Skyrmion}

To compute the angular momentum carried by a skyrmion in the system, we consider the angular momentum density due to the electromagnetic field.
For a  static spin texture it  is clearly zero, because $\vec{E}=0$. For a time dependent texture it is little harder to prove. Consider, \begin{eqnarray}
N_x(r,\theta,t) &=& \sin \alpha(r,t) \cos \beta(\theta,t), \nonumber \\
N_y(r,\theta,t) &=& \sin \alpha(r,t) \sin \beta(\theta,t), \nonumber \\
N_z(r,t) &=& \cos \alpha(r,t), \nonumber
\end{eqnarray}
where $\alpha(r,t)$ and $\beta(\theta,t)$ are smooth functions, and $\alpha(r,t)$ satisfies the boundary conditions $\alpha(r=0,t) = 0$ and $\alpha(r\rightarrow \infty,t) = \pi$, for any $t$, 
and $\frac{\partial \alpha(r,t)}{\partial r}|_{r\rightarrow\infty} = \frac{\partial \alpha(r,t)}{\partial t}|_{r\rightarrow\infty}=0$ in the far field limit. The unitary matrix is now time dependent.
After a little algebra, we obtain the time-dependent gauge fields in the far field limit to be 
\begin{eqnarray}
A_{x}^{s}(r,\theta,t)&=& \frac{-2 \sin\theta}{r} \frac{\partial \beta(\theta,t)}{\partial \theta}, \\
A_{y}^{s}(r,\theta,t)&=& \frac{2 \cos\theta}{r} \frac{\partial \beta(\theta,t)}{\partial \theta}.
\end{eqnarray}
So, $\Phi(\theta,t)= A_{t}^{s}(\theta,t)= 2 \frac{\partial \beta(\theta,t)}{\partial t}$ and 
$\vec{A}^{s}(r,\theta,t) = A_{x}^{s}(r,\theta,t)\hat{x} + A_{y}^{s}(r,\theta,t)\hat{y} = A_{\theta}^{s}(r,\theta,t) \hat{\theta}$,
where
\begin{equation} 
A_{\theta}^{s}(r,\theta,t) = \frac{2}{r} \frac{\partial \beta(\theta,t)}{\partial \theta}.
\end{equation}
Therefore, the electric field will have a non-zero $\hat{\theta}-$component, $\vec{E} = E_{\theta} \hat{\theta}$, and the magnetic field will have a non-zero $\hat{z}-$component, $\vec{B} = B_{z} \hat{z}$, 
where 
\begin{eqnarray}
E_{\theta} &=& - \frac{1}{r} \frac{\partial A^{s}_{t}(\theta,t) }{\partial \theta} - \frac{\partial A^{s}_{\theta}(r,t) }{\partial t} = - \frac{4}{r} \frac{\partial^{2} \beta(\theta,t) }{\partial \theta \partial t} \\
B_{z} &=& \frac{\partial A^{s}_{\theta}(r,t) }{\partial r}  = - \frac{2}{r^2} \frac{\partial  \beta(\theta,t) }{\partial \theta}.
\end{eqnarray}
As a result, the angular momentum density still vanishes,
\begin{equation}
\vec{L}_{\text{field}}= \frac{1}{4\pi c} \vec{r} \times (E_{\theta} \hat{\theta} \times B_{z} \hat{z})=0.
\end{equation}
It is possible that  superconductivity with non-zero angular momentum may be realized when the interaction between skyrmions is included, but we do not know how to prove it.  It would be
interesting to explore what other kinds of quantum numbers are carried by the
topological textures in the model we have studied.

\section{\label{Sec:pumping}Quantized charge pumping}

In Sec.\ref{Sec:charge}, we considered a static spin texture and obtained charge-$2e$ skyrmions in the system. If we consider a time-dependent spin texture, which has a slow variation in one spatial direction, say, $\hat{y}$, and is uniform in the other, $\hat{x}$, charge will be pumped from one side of the system to the other  along $\hat{x}$.~\cite{Yang:2011} This charge pumping effect can be understood from the effective gauge action, which is
\begin{equation}
S_{\text{eff}}[A^{c}_{\mu},A^{s}_{\mu}]=\frac{C_{2}}{4\pi} \int d^3\it{x} \epsilon^{\mu\nu\lambda}A^c_{\mu}\partial_{\nu}A^s_{\lambda},
\end{equation}
where the integral is over the real time, $t$, instead of the imaginary time, $\tau$. Therefore, the charge current induced by the spin gauge field will be
\begin{eqnarray}
j^c_{\mu} &=& \frac{\delta S_{\text{eff}}[A^{c}_{\mu},A^{s}_{\mu}]}{\delta A^{c}_{\mu}} \nonumber \\
 &=& \frac{C_{2}}{4\pi} \epsilon^{\mu\nu\lambda}\partial_{\nu}A^s_{\lambda} \nonumber \\
 &=& \frac{C_{2}}{8\pi} \epsilon^{\mu\nu\lambda}F^{s}_{\nu\lambda},
\end{eqnarray}
where we define the spin gauge flux $F^{s}_{\mu\nu} \equiv \partial_{\mu}A^{s}_{\nu}-\partial_{\nu}A^{s}_{\mu}$. After some straightforward algebra (see Appendix \ref{Sec:flux}), the spin gauge flux can be written in terms of the $\hat{N}$-vector,
\begin{equation}
F^{s}_{\mu\nu} 
= \hat{N} \cdot [(\partial_{\mu}\hat{N})\times(\partial_{\nu}\hat{N})].
\end{equation}
As a result, even in absence of the external electromagnetic field, a charge current may be induced by a time-dependent inhomogeneous spin texture because 
\begin{equation}
j^c_{\mu} = \frac{C_{2}}{8\pi} \epsilon^{\mu\nu\lambda} \hat{N} \cdot [(\partial_{\nu}\hat{N})\times(\partial_{\lambda}\hat{N})].
\end{equation}

To demonstrate the charge response induced by the spin texture, we consider the following configuration with unit Pontryagin index,
\begin{equation}
\hat{N}(y,t) =\left[ \sin \theta(t) \cos \phi(y), \sin \theta(t) \sin \phi(y), \cos \theta(t) \right],
\end{equation}
where $\theta(t)$ and $\phi(y)$ are smooth functions of $t$ and $y$, respectively, with boundary conditions $\theta(t=0)=0$, $\theta(t=\tau_p)=\pi$, and $\phi(y\rightarrow \pm \infty)=\pm \pi$. Therefore, we have an induced charge current along the $\hat{x}$-direction,
\begin{eqnarray}
j^c_{x} &=& \frac{C_{2}}{8\pi} \epsilon^{x\nu\lambda} \hat{N} \cdot [(\partial_{\nu}\hat{N})\times(\partial_{\lambda}\hat{N})] \nonumber\\
&=& \frac{C_{2}}{4\pi}\hat{N} \cdot [(\partial_{y}\hat{N})\times(\partial_{t}\hat{N})].
\end{eqnarray}
Interestingly, we can show that the pumped charge is quantized,
\begin{eqnarray}
Q_{\text{pumped}} &=& \int^{\tau_p}_{0} dt \int^{\infty}_{-\infty} dy \; j^c_{x} \nonumber\\
&=& \frac{C_{2}}{4\pi} \int^{\tau_p}_{0} dt \int^{\infty}_{-\infty} dy \; 
\hat{N} \cdot [(\partial_{y}\hat{N})\times(\partial_{t}\hat{N})] \nonumber\\
&=& \frac{C_{2}}{4\pi} \int^{\pi}_{0} d\theta \int^{\pi}_{-\pi} d\phi \; 
\hat{N} \cdot [(\partial_{\theta}\hat{N})\times(\partial_{\phi}\hat{N})] \nonumber\\
&=& C_{2},
\end{eqnarray}
where we have used that, for the spin texture with unit Pontryagin index, 
\begin{equation}
 \int^{\pi}_{0} d\theta \int^{\pi}_{-\pi} d\phi \; 
\hat{N} \cdot [(\partial_{\theta}\hat{N})\times(\partial_{\phi}\hat{N})] =4\pi.
\end{equation}
After restoring the unit of charge, we have $Q_{\text{pumped}} = C_{2} \, e$. So far we have considered the spin texture with unit Pontryagin index. If the spin texture is generalized to a general Pontryagin index, $N_{P}$, then the pumped charge will be $Q_{\text{pumped}} = C_{2} N_{P} \, e$.

How could we observe this charge pumping  experimentally? 
We need to control the direction of the $\hat{N}$-vector so that it can be the time-dependent inhomogeneous spin texture discussed above. In topological chiral magnets,~\cite{Yang:2011} the $\hat{N}$-vector is the net ferromagnetic moment, which aligns along the external magnetic field, so one can apply a time-dependent magnetic field $\vec{H}(t)=H(t)\hat{x}$ coupling to the $\hat{N}$-vector and control the magnitude of $\hat{x}$-component of $\hat{N}$. 

In the mixed triplet $d$-density wave, however, the situation is more complicated. In the presence of an external magnetic field, there will be a spin flop transition  and the $\hat{N}$-vector will lie in the plane perpendicular to the external field.~\cite{Nersesyan:1991} In other words, we cannot fully control the direction of $\hat{N}$ with a time-dependent magnetic field. Therefore, it would be a challenge to measure the pumped charges in the system.

Nevertheless, the charge pumping effect provides, at least,  a different conceptual approach to probe the topological properties of the system in addition to the quantized spin Hall conductance. For the quantum spin Hall effect, a spin current is induced by the external electric field,~\cite{Hsu:2011} whereas for the charge pumping effect, a charge current will be induced by the spin texture. It would, of course, be interesting if one can manipulate the $\hat{N}$-vector experimentally because the charge current is easier to detect than the spin current.

\section{Discussion and application to the hidden order state in $\mathrm{URu_{2}Si_{2}}$}
\label{discussion}
 
 There are two  points that we have glossed over. The first is rather simple: in the ordered phase at $T=0$, there are also Goldstone modes that can be easily seen by integrating out the fermions resulting in a 
 non-linear $\sigma$-model involving $\hat{N}$, the form of which is entirely determined by symmetry. These do not lead to any interesting physics, such as charge-$2e$ skyrmions that condense into a superconducting state. 
 At finite temperatures they could lead to a renormalized classical behavior.~\cite{Chakravarty:1989} The second point is more subtle: we have assumed that the hedgehog configurations are absent. This would require, as pointed out by Grover and Senthil~\cite{Grover:2008}, that the energy of the Skyrmion (especially in the limit $\Delta_{0}\to 0$) is smaller than individual pairs of electrons, a question that is likely to be model dependent. If this assumption is correct, however, the transition from the the mixed  $d$-density wave state to the superconducting state will correspond to a deconfined quantum critical point, which otherwise would have been a first order transition, as in Landau theory.~\cite{Senthil:2004,*Senthil:2004b}

We suggest that the  superconducting phase driven by the skyrmion condensate may be realized in the URu$_2$Si$_2$, which hosts an exotic hidden order (HO) phase, with broken translatenal symmetry  below $T_{HO}\approx 17.5K$ and a superconducting phase be\text{low} $T_c \approx 1.5K$.~\cite{Mydosh:2011} Recently, Fujimoto~\cite{Fujimoto:2011} proposed  a triplet $d$-density wave  with the order parameter $\langle c_{k,1,\alpha}^{\dagger} c_{k+Q_0,2,\beta}\rangle =  \vec{d}(k) \cdot \vec{\sigma}_{\alpha\beta}$ with  $\vec{d}(k)=i(\Delta_1 \sin \frac{(k_x-k_y)}{\sqrt{2}} \sin k_z,0,0)$ to describe this state;~\cite{Fujimoto:2011} here $1$ and $2$ refer to  two different bands and $Q_{0}=(0,0,1)$ is the  nesting vector; even the  earlier work in Ref.~\onlinecite{Ikeda:1998} involving circulating spin current is not entirely unrelated. The order parameter considered in Ref.~\onlinecite{Fujimoto:2011}  is different but a close cousin of the order parameter considered in our work; the circulating staggered spin currents in Ref.~\onlinecite{Fujimoto:2011} lie on the diagonal planes instead and the crucial $d_{xy}$ part is missing there. That the currents are in the diagonal planes instead of being square planar is conceptually not important, but is necessary to explain the nematicity observed in the experiments.~\cite{Okazaki:2011}  We now discuss the role of spin-orbit coupling before making our final comments.
\subsection*{Spin-orbit coupling}
 It will be shown below that the order of magnitude of the spin-orbit energy $E_{\text{SO}}\approx [(\hat{N}\cdot \hat{z})^{2}-1] (\Lambda_{0}^{2}/W)(W_{0}/W)^{2}[1+{\cal O}(W_{0}/W)^{2}]$, correcting a mistake in Ref.~\onlinecite{Nersesyan:1991}. Here $\Lambda_{0}$ is the strength of the spin-orbit coupling, given by 
 \begin{equation}
 \mathcal{H}_{\text{SO}}=\sum_{k} c^{\dagger}_{k\alpha}\vec{\Lambda}(k)\cdot \vec{\sigma}_{\alpha\beta} c_{k\beta}, 
 \end{equation}
where  $\vec{\Lambda}(k) = (\Lambda_{0}/\sqrt{2})[\hat{x}  \sin k_{y}-\hat{y} \sin k_{x} ]$.
In the presence of spin-orbit coupling, the Hamiltonian is
\begin{widetext}
\begin{eqnarray}
&&\mathcal{H}_{\text{total}}=\mathcal{H}+\mathcal{H}_{SO} \nonumber \\
&=&\sum_{k} \Psi_{k}^{\dagger}   \left(
  \begin{array}{cccc}
   \epsilon_{k}  & \Delta_k+i N_z W_k      & \Lambda_x(k)-i\Lambda_y(k)               & i W_k (N_x-iN_y)            \\
   \Delta_k-i N_z W_k     & -\epsilon_{k} & -i W_k (N_x-iN_y)                 & -\Lambda_x(k)+i\Lambda_y(k)             \\
   \Lambda_x(k)+i\Lambda_y(k)                 & i W_k (N_x+iN_y)                  & \epsilon_{k}  & \Delta_k-i N_z W_k \\
 -  i W_k (N_x+iN_y)                 & -\Lambda_x(k)-i\Lambda_y(k)                  & \Delta_k+i N_z W_k     & -\epsilon_{k}
  \end{array}
\right)    \Psi_k,
\end{eqnarray}
 where   $\Psi_{k}^{\dagger}$ is the four-component spinor $(c_{k,\uparrow}^{\dagger}, c_{k+Q,\uparrow}^{\dagger}, c_{k,\downarrow}^{\dagger}, c_{k+Q,\downarrow}^{\dagger})$.
In the absence of spin-orbit coupling, the eigenvalues are $\pm E_{0k}$ with $ E_{0k}=\sqrt{\epsilon_k^2 + W_k^2 +\Delta_k^2}$. On the other hand when  spin-orbit coupling is present, the eigenvalues of the upper and lower bands now become $\lambda_{\text{up},\pm}=E_{k,\pm}$,  $\lambda_{\text{low},\pm}=-E_{k,\pm}$, respectively, where 
\begin{equation}
 E_{k,\pm}=\sqrt{\epsilon_k^2 + W_k^2 +\Delta_k^2+\Lambda_k^2 \pm 2 \left[ (\epsilon_k^2 + W_k^2) \Lambda_k^2  - W_k^2(\hat{N} \cdot \vec{\Lambda}_k)^2
\right]^{\frac{1}{2}} }
\end{equation}
with $\Lambda_k^2 \equiv |\vec{\Lambda}_k|^2 = \Lambda_x^2(k)+\Lambda_y^2(k)$. When the $d_{xy}$ component is absent,  $\Delta_k=0$, and the results of Ref.~\onlinecite{Nersesyan:1991} are recovered. Consider the following two cases separately.
\end{widetext}

\subsubsection{$\hat{N}\parallel\hat{z}$}
Since the chemical potential is at the mid-gap, we can focus on the lower bands.
When $\hat{N}=\hat{z}$, we have $\hat{N} \cdot \vec{\Lambda}_k=0$ and
\begin{eqnarray}
 \lambda^{z}_{\text{low},\pm}&=&-\sqrt{E_{0k}^2+\Lambda_k^2 \pm 2 \left[E_{0k}^2 \Lambda_k^2 \right]^{\frac{1}{2}} } \nonumber\\
&=&-E_{0k} \mp |\vec{\Lambda}_k|
\end{eqnarray}
Assuming that $\Lambda_0 \ll W_0, \Delta_0 \ll W$ with the electronic bandwidth $W=8t$, the change in the ground state energy will be
\begin{eqnarray}
E_{\text{SO}} &=& \sum_k \left[ (\lambda^{z}_{\text{low},+}+\lambda^{z}_{\text{low},-})-2 (-E_{0k}) \right] \nonumber\\
&=& \sum_k \left[ (-E_{0k} - |\vec{\Lambda}_k|-E_{0k} + |\vec{\Lambda}_k|)+2 E_{0k} \right] \nonumber\\
&=& 0
\end{eqnarray}

\subsubsection{$\hat{N}\perp \hat{z}$}

When $\hat{N}$ lies in $xy$-plane, we have $\hat{N} \cdot \vec{\Lambda}_k=|\vec{\Lambda}_k| \cos \phi_k$, 
where $\phi_k$ is the angle between $\hat{N}$ and $\vec{\Lambda}_k$, and
\begin{equation}
\cos \phi_k = \frac{\hat{N} \cdot \vec{\Lambda}_k}{|\vec{\Lambda}_k|}=\frac{N_x\Lambda_x(k)+N_y\Lambda_y(k)}{\sqrt{\Lambda_x^2(k)+\Lambda_y^2(k)}}.
\end{equation}
\begin{widetext}
The eigenvalues of the lower bands are now
\begin{eqnarray}
\lambda^{xy}_{\text{low},\pm}&=&-\sqrt{E_{0k}^2+\Lambda_k^2 \pm 2 \left[E_{0k}^2 \Lambda_k^2  - W_k^2\Lambda_k^2 \cos^2 \phi_k \right]^{\frac{1}{2}} } \nonumber\\
&\approx& -E_{0k} 
\mp (1-\frac{1}{2} \frac{W_k^2}{E_{0k}^2} ) |\vec{\Lambda}_k|
- \frac{1}{2} \frac{W_k^2}{E_{0k}}  
\frac{\Lambda_k^2}{E_{0k}^2} (1+\mathcal{O}(\frac{W_k^2}{E_{0k}^2})), 
\end{eqnarray}
where we have used $\cos^2 \phi_{k} \approx \mathcal{O}(1)$. Notice that the signs of the second order terms for $\lambda^{xy}_{\text{low},+}$ and $\lambda^{xy}_{\text{low},-}$ are both negative, leading to the net change in the ground state energy, which is opposite to the $\hat{N}=\hat{z}$ case. 
Assuming that $\Lambda_0 \ll W_0, \Delta_0 \ll W$, the change in the ground state energy per lattice site will be
\begin{eqnarray}
E_{\text{SO}}&=& \sum_k \left[ (\lambda^{xy}_{\text{low},+}+\lambda^{xy}_{\text{low},-})-2 (-E_{0k} ) \right] \nonumber\\
&\approx& - \sum_k  \frac{\Lambda_k^2 W_k^2}{E_{0k}^3} \left[ 1+\mathcal{O}(\frac{W_k^2}{E_{0k}^2}) \right] \nonumber\\
&=&  -   \frac{\Lambda_0^2}{W} \left( \frac{W_0}{W}\right)^{2}\left[1+\mathcal{O}\left(\frac{W_0}{W}\right)^{2} \right]  < 0,
\end{eqnarray}
 Therefore, $\hat{N}$-vector should  lie in the $xy$-plane in the presence of spin-orbit interaction and the result stated above follows.
 \end{widetext}
 As large as the spin-orbit coupling may be for U atoms,  $E_{\text{SO}}$ is  still a small energy scale. However, if other anisotropies are absent, the order parameter would be in the $XY$-plane, resulting in vortices;  exchange anisotropy can also result in an easy-axis anisotropy, in which case spin textures could be Ising domain walls that can trap electrons. Although skyrmions are finite energy solutions, vortices cost infinite energy unless they are bound in pairs. We speculate that charge $2e$-skyrmionic condensation is a more likely scenario, but the crossover in the texture is an interesting topic for further research.

The following remarks about  URu$_2$Si$_2$  are relevant:
in both  magnetic field-temperature ($H-T$)  and pressure-temperature ($P-T$)  phase diagrams, the superconducting phase is enclosed within the HO phase.~\cite{Mydosh:2011} It implies that the superconducting phase is closely related to the HO phase, and is probably induced by it. 
Throughout our calculation, ignoring of course skyrmions, we have assumed  that the system is half-filled. The lower band is filled and the upper band is empty, and the topological invariant is quantized. If this is not the case, then there will be no quantized spin Hall conductance, but an induced superconducting phase from charge $2e$-skyrmionic condensation;  doping will result in conducting mid-gap states, as in polyacetylene.~\cite{Su:1988} Of course, such a topological superconducting phase is very sensitive to disorder. Indeed, this may be supported by the destruction of the HO and SC phases with $4\%$ Rh substitution on the Ru site.~\cite{Mydosh:2011}
To summarize, we can find a rationale for a hidden order phase enclosing a superconducting phase at \text{low}er temperatures. 

\begin{acknowledgments}
This work is supported by NSF under Grant No. DMR-1004520.  We are grateful to E. Abrahams,  E. Fradkin, S. Kivelson for useful  comments regarding the manuscript. Special thanks are due to S. Raghu for his continued
interest in our work and for collaboration at  earlier stages.  Liang Fu, Tarun Grover, and Igor Herbut   have made important suggestions. H. Y. Kee gave us some confidence in regard to the applicability of our ideas to the hidden order state in $\mathrm{URu_{2}Si_{2}}$. S. C. also acknowledge support from NSF Grant No. PHY-1066293 and the  hospitality of the Aspen Center for Physics where the work germinated. 
\end{acknowledgments}

\appendix
\section{\label{Sec:NLSM}Derivation of the non-linear $\sigma$-model}

To derive the non-linear $\sigma$-model, we compute the effective action by integrating out fermions. We start with the action $S = \sum_{j=1}^{3} S_j$, where
\begin{eqnarray}
S_j \equiv \int d^3 x \bar{\psi}_{j} \left[ G_{j}^{-1} \right] \psi_{j},
\end{eqnarray}
with $G_{j}^{-1} \equiv G_{0,j}^{-1} + \Sigma_j$.

For $j=1,2$, we have
\begin{eqnarray}
G_{0,j}^{-1} &\equiv& -i \sigma^0\tau^z \partial_{\tau}
- 2it \sigma^0 \tau^{y} (\eta_{j}\partial_{x}+\partial_{y}), \label{Eq:G12} \\
\Sigma_j  &\equiv& i \eta_{j} \Delta_0 \sigma^0 \tau^{0}  
- i\frac{W_0}{2} (\vec{\sigma} \cdot \hat{N}) \tau^x 
(-\eta_{j}\partial_{x}+\partial_{y}), 
\end{eqnarray}
and for $j=3$, we have
\begin{eqnarray}
G_{0,3}^{-1} &\equiv& -i \sigma^0 \tau^z \partial_{\tau}, \label{Eq:G3} \\
\Sigma_3  &\equiv& W_0 (\vec{\sigma} \cdot \hat{N}) \tau^x.
\end{eqnarray}

The effective action will be $S_{eff} = \sum_{j=1}^{3} S_{j,eff}$ with
\begin{eqnarray}
S_{j,eff} &=& - \ln \left[  \int D\bar{\psi}_{j} D\psi_{j} e^{-S_j} \right] \nonumber \\
&=& - \ln \left[ \det|G_j^{-1}| \right],
\end{eqnarray}
where the fermion operators can be integrated out easily since the Hamiltonian has only bilinear fermion operator terms. Using the mathematical identity $ \ln \det |A| = \textrm{tr} \ln A$ with $\textrm{tr}$ being the trace, we have
\begin{widetext}
\begin{eqnarray}
S_{j,eff} 
&=& - \textrm{tr} \ln G_{0,j}^{-1} \left[  1 + G_{0,j} \Sigma_j \right]  \nonumber \\
&=& - \textrm{tr} \ln G_{0,j}^{-1} - \textrm{tr} \left[ G_{0,j} \Sigma_j \right] + \frac{1}{2} \textrm{tr} \left[ G_{0,j} \Sigma_j G_{0,j} \Sigma_j \right] +  \cdots,
\end{eqnarray}
\end{widetext}
where we have used $\ln (1+x) = x -\frac{x^2}{2} + \cdots$.

The zeroth order term is the effective action for free particles and the first order term vanishes, so our goal is to compute the second order terms:
\begin{widetext}
\begin{eqnarray}
S_{j,eff}^{(2)} &\equiv& \frac{1}{2} \textrm{tr} \left[ G_{0,j} \Sigma_j G_{0,j} \Sigma_j \right] \nonumber \\
&=& \frac{1}{2} \int d\tau \int d\tau^{\prime} \int d^2 x \int d^2 x^{\prime} \textrm{Tr} \left[ G_{0,j}(x,\tau;x^{\prime},\tau^{\prime} ) \Sigma_j(x^{\prime},\tau^{\prime}) G_{0,j}(x^{\prime},\tau^{\prime};x,\tau) \Sigma_j(x,\tau) \right] \nonumber\\
 &=& \frac{1}{2} \sum_{\tilde{k}, \tilde{q}}  \textrm{Tr} \left[ G_{0,j}(\tilde{k}) \Sigma_j(\tilde{q}) G_{0,j}(\tilde{k}+\tilde{q}) \Sigma_j(-\tilde{q}) \right], 
\end{eqnarray}
\end{widetext}
where $\tilde{k}\equiv(k_0,k_x,k_y)$, $\tilde{q} \equiv(q_0,q_x,q_y)$, and $G_{0,j}(\tilde{k})$ can be obtained by inverting Eq.(\ref{Eq:G12}) and Eq.(\ref{Eq:G3}).  

Putting all together, taking long wavelength limit ($\tilde{q} \rightarrow 0$) and keeping only terms up to the second order derivative, we have, for $j=1,2$,
\begin{widetext}
\begin{eqnarray}
S_{j,eff}^{(2)} 
&\approx&
 2 \sum_{\tilde{k}, \tilde{q}} \frac{1}{k_0^2 + 4t^2(\eta_{j}k_x+k_y)^2 } 
\left[ -\Delta_0^2 +(\frac{W_0}{2})^2 (-\eta_{j}q_x+q_y)^2 ( \hat{N}_{\tilde{q}} \cdot \hat{N}_{-\tilde{q}})\right],
\end{eqnarray}
\end{widetext}
where terms which are odd in $\tilde{k}$ and $\tilde{q}$ are dropped. 

Using the relation $\sum_{\tilde{q}} f_{\tilde{q}} f_{-\tilde{q}} = \int d \tau d^2x |f(\vec{x},\tau)|^2$, we obtain
\begin{eqnarray}
S_{1,eff}^{(2)} + S_{2,eff}^{(2)} 
&\approx& \frac{1}{g_1} \int d \tau d^2x  \left[  \left| \partial_X \hat{N} \right|^2 + \left| \partial_Y \hat{N} \right|^2 \right], \nonumber\\
\end{eqnarray}
where the constant terms are dropped, $(X,Y)$ is the coordinate after a $\pi/4$ rotation, and 
\begin{eqnarray}
\frac{1}{g_1} &\equiv& \sum_{\tilde{k}} \frac{-W_0^2}{2(k_0^2 + 4t^2(k_x+k_y)^2)}
\end{eqnarray}

Similarly, for $j=3$, we obtain
\begin{eqnarray}
S_{3,eff}^{(2)} 
&\approx& -2 \sum_{\tilde{k}, \tilde{q}}  \frac{W_0^2}{k_0^2} (\frac{q_0}{k_0})^2  (\hat{N}_{\tilde{q}} \cdot \hat{N}_{-\tilde{q}}) \nonumber\\
&=& \frac{1}{g_3} \int d \tau d^2x  \left|\partial_{\tau} \hat{N} \right|^2,
\end{eqnarray}
where 
\begin{eqnarray}
\frac{1}{g_3} &\equiv& \sum_{\tilde{k}} \frac{2 W_0^2}{k_0^4}.
\end{eqnarray}

Therefore, we obtain the non-linear sigma model,
\begin{eqnarray}
S_{eff} 
&\approx& \frac{1}{g} \int d \tau d^2x \left|\partial_{\mu} \hat{N}\right|^2,
\end{eqnarray}
where the constant terms and higher order terms are dropped, and it is rescaled in order to obtain a familiar form.

\section{\label{Sec:Chern}Chern-Simons coefficients}
In this appendix we are going to prove that
\begin{eqnarray}
{\cal N}(G_{\sigma})&=&\frac{\epsilon_{\mu\nu\lambda}}{24\pi^2} \text{Tr} \left[ \int d^3k 
G_{\sigma} \frac{\partial G_{\sigma}^{-1}}{\partial k_{\mu}} 
G_{\sigma} \frac{\partial G_{\sigma}^{-1}}{\partial k_{\nu}}
G_{\sigma} \frac{\partial G_{\sigma}^{-1}}{\partial k_{\lambda}} \right] \nonumber \\
&=&
-\int \frac{d^2k}{4\pi} 
\hat{h}_{\sigma} \cdot
\frac{\partial \hat{h}_{\sigma}}{\partial k_{x}} \times 
\frac{\partial \hat{h}_{\sigma}}{\partial k_{y}}.
\end{eqnarray}

We start by taking $(\mu,\nu,\lambda)$ to be $(0,x,y)$, and obtain 
\begin{eqnarray}
G_{\sigma}\frac{\partial G_{\sigma}^{-1}}{\partial \omega}
&=& \frac{1}{(i\omega)^2-|\vec{h}_{\sigma}|^2} \left[ 
(i \omega \hat{I} + \hat{\tau} \cdot \vec{h}_{\sigma})
\cdot
(i \hat{I}) 
\right]  \nonumber \\
&=& \frac{1}{(i\omega)^2-|\vec{h}_{\sigma}|^2} 
(- \omega \hat{I} + i \hat{\tau} \cdot \vec{h}_{\sigma}),
\end{eqnarray}
and
\begin{widetext}
\begin{eqnarray}
G_{\sigma}\frac{\partial G_{\sigma}^{-1}}{\partial k_{x}} &=& \frac{1}{(i\omega)^2-|\vec{h}_{\sigma}|^2}
\left[ i \omega \hat{I} + \hat{\tau} \cdot \vec{h}_{\sigma} \right] (- \hat{\tau} \cdot 
\frac{\partial \vec{h}_{\sigma}}{\partial k_{x}} )\nonumber \\
&=& \frac{-1}{(i\omega)^2-|\vec{h}_{\sigma}|^2} \left[ 
(\vec{h}_{\sigma} \cdot \frac{\partial \vec{h}_{\sigma}}{\partial k_{x}}) \hat{I} 
+i  \hat{\tau} \cdot ( 
\omega \frac{\partial \vec{h}_{\sigma}}{\partial k_{x}} 
+ \vec{h}_{\sigma} \times \frac{\partial \vec{h}_{\sigma}}{\partial k_{x}} )
\right],
\end{eqnarray}
where we have used the matrix identity 
$ (\hat{\tau} \cdot \vec{a})(\hat{\tau} \cdot \vec{b} ) 
= (\vec{a} \cdot \vec{b}) \hat{I} + i \hat{\tau} \cdot (\vec{a} \times \vec{b}) $.
Similarly,
\begin{eqnarray}
G_{\sigma}\frac{\partial G_{\sigma}^{-1}}{\partial k_{y}}
&=& \frac{-1}{(i\omega)^2-|\vec{h}_{\sigma}|^2} \left[ 
(\vec{h}_{\sigma} \cdot \frac{\partial \vec{h}_{\sigma}}{\partial k_{y}}) \hat{I} 
+i  \hat{\tau} \cdot ( 
\omega \frac{\partial \vec{h}_{\sigma}}{\partial k_{y}} 
+ \vec{h}_{\sigma} \times \frac{\partial \vec{h}_{\sigma}}{\partial k_{y}} )
\right].
\end{eqnarray}

Therefore,
\begin{eqnarray}
G_{\sigma} \frac{\partial G_{\sigma}^{-1}}{\partial k_{x}}
G_{\sigma} \frac{\partial G_{\sigma}^{-1}}{\partial k_{y}} 
&=& \frac{1}{((i\omega)^2-|\vec{h}_{\sigma}|^2)^2} 
\left\{ 
(\vec{h}_{\sigma} \cdot \frac{\partial \vec{h}_{\sigma}}{\partial k_{x}})
(\vec{h}_{\sigma} \cdot \frac{\partial \vec{h}_{\sigma}}{\partial k_{y}}) \hat{I} \right. \nonumber \\
&&+ i  \hat{\tau}\cdot \left[
(\vec{h}_{\sigma} \cdot \frac{\partial \vec{h}_{\sigma}}{\partial k_{x}})
( \omega \frac{\partial \vec{h}_{\sigma}}{\partial k_{y}} 
+ \vec{h}_{\sigma} \times \frac{\partial \vec{h}_{\sigma}}{\partial k_{y}} )
+ (\vec{h}_{\sigma} \cdot \frac{\partial \vec{h}_{\sigma}}{\partial k_{y}}) 
( \omega \frac{\partial \vec{h}_{\sigma}}{\partial k_{x}} 
+ \vec{h}_{\sigma} \times \frac{\partial \vec{h}_{\sigma}}{\partial k_{x}} )
\right] \nonumber \\
&& \left. - \left[
\hat{\tau} \cdot ( 
\omega \frac{\partial \vec{h}_{\sigma}}{\partial k_{x}} 
+ \vec{h}_{\sigma} \times \frac{\partial \vec{h}_{\sigma}}{\partial k_{x}} )
\right]
\left[
\hat{\tau} \cdot ( 
\omega \frac{\partial \vec{h}_{\sigma}}{\partial k_{y}} 
+ \vec{h}_{\sigma} \times \frac{\partial \vec{h}_{\sigma}}{\partial k_{y}} )
\right]
\right\}.
\end{eqnarray}
\end{widetext}

Since we are going to multiply it with the antisymmetric tensor $\epsilon_{\mu\nu\lambda}$, the terms which are symmetric under $(x \leftrightarrow y)$ will vanish. Therefore, only the last term in the bracket contributes, 
\begin{widetext}
\begin{eqnarray}
&&\left[
\hat{\tau} \cdot ( 
\omega \frac{\partial \vec{h}_{\sigma}}{\partial k_{x}} 
+ \vec{h}_{\sigma} \times \frac{\partial \vec{h}_{\sigma}}{\partial k_{x}} )
\right]
\left[
\hat{\tau} \cdot 
( \omega \frac{\partial \vec{h}_{\sigma}}{\partial k_{y}} 
+ \vec{h}_{\sigma} \times \frac{\partial \vec{h}_{\sigma}}{\partial k_{y}} )
\right]  \nonumber\\
&=&  i \hat{\tau} \cdot \left[
\omega^2(\frac{\partial \vec{h}_{\sigma}}{\partial k_{x}} \times \frac{\partial \vec{h}_{\sigma}}{\partial k_{y}})  + \omega \vec{h}_{\sigma} (\frac{\partial \vec{h}_{\sigma}}{\partial k_{x}} \cdot \frac{\partial \vec{h}_{\sigma}}{\partial k_{y}})
- \omega \frac{\partial \vec{h}_{\sigma}}{\partial k_{y}}
( \vec{h}_{\sigma} \cdot \frac{\partial \vec{h}_{\sigma}}{\partial k_{x}} )
 \right. \nonumber \\
&& 
- \omega \vec{h}_{\sigma} (\frac{\partial \vec{h}_{\sigma}}{\partial k_{x}} \cdot \frac{\partial \vec{h}_{\sigma}}{\partial k_{y}})
+ \omega \frac{\partial \vec{h}_{\sigma}}{\partial k_{x}}
( \vec{h}_{\sigma} \cdot \frac{\partial \vec{h}_{\sigma}}{\partial k_{y}} )
+ 
\left. 
(\vec{h}_{\sigma} \cdot \frac{\partial \vec{h}_{\sigma}}{\partial k_{x}} 
 \times \frac{\partial \vec{h}_{\sigma}}{\partial k_{y}}) \vec{h}_{\sigma} 
\right],
\end{eqnarray}
\end{widetext}
where we used the following mathematical identities:
\begin{eqnarray*}
\vec{a} \times (\vec{b} \times \vec{c}) &=& 
\vec{b}(\vec{a}\cdot\vec{c}) - \vec{c}(\vec{a}\cdot\vec{b}), \\
(\vec{a} \times \vec{b}) \times (\vec{a} \times \vec{c}) &=&
(\vec{a} \cdot (\vec{b} \times \vec{c}))\vec{a}.
\end{eqnarray*}

Therefore, after combining with $\epsilon_{0xy}$ and taking the trace, we have
\begin{widetext}
\begin{eqnarray}
&&\epsilon_{0xy} \textrm{Tr} \left[ G_{\sigma}\frac{\partial G_{\sigma}^{-1}}{\partial \omega} 
G_{\sigma} \frac{\partial G_{\sigma}^{-1}}{\partial k_{x}}
G_{\sigma} \frac{\partial G_{\sigma}^{-1}}{\partial k_{y}} \right]\nonumber \\
&=& \frac{-1}{((i\omega)^2-|\vec{h}_{\sigma}|^2)^3} \textrm{Tr} 
\left\{
-i \omega\hat{\tau} \cdot
\left[ 
\omega^2(\frac{\partial \vec{h}_{\sigma}}{\partial k_{x}} \times \frac{\partial \vec{h}_{\sigma}}{\partial k_{y}})
+(\vec{h}_{\sigma} \cdot \frac{\partial \vec{h}_{\sigma}}{\partial k_{x}} 
 \times \frac{\partial \vec{h}_{\sigma}}{\partial k_{y}}) \vec{h}_{\sigma} 
\right] 
\right. \nonumber \\
&& \hspace{1in} \left. - (\hat{\tau} \cdot \vec{h}_{\sigma})
\left[\hat{\tau} \cdot 
[
\omega^2(\frac{\partial \vec{h}_{\sigma}}{\partial k_{x}} \times \frac{\partial \vec{h}_{\sigma}}{\partial k_{y}})
+(\vec{h}_{\sigma} \cdot \frac{\partial \vec{h}_{\sigma}}{\partial k_{x}} 
 \times \frac{\partial \vec{h}_{\sigma}}{\partial k_{y}}) \vec{h}_{\sigma} 
]
\right]
\right\} \nonumber \\
&=&
\frac{2}{((i\omega)^2-|\vec{h}_{\sigma}|^2)^3} 
 \left[
\vec{h}_{\sigma}
\cdot
[\omega^2(\frac{\partial \vec{h}_{\sigma}}{\partial k_{x}} \times \frac{\partial \vec{h}_{\sigma}}{\partial k_{y}})
+(\vec{h}_{\sigma} \cdot \frac{\partial \vec{h}_{\sigma}}{\partial k_{x}} 
 \times \frac{\partial \vec{h}_{\sigma}}{\partial k_{y}}) \vec{h}_{\sigma} ]
\right]   \nonumber \\
&=&
\frac{-2}{((i\omega)^2-|\vec{h}_{\sigma}|^2)^2} 
(\vec{h}_{\sigma} \cdot \frac{\partial \vec{h}_{\sigma}}{\partial k_{x}} 
 \times \frac{\partial \vec{h}_{\sigma}}{\partial k_{y}}), 
\end{eqnarray}
\end{widetext}
where we have used the fact that Pauli matrices are traceless, so the only contribution will be the term proportional to $\hat{I}$.

We have six non-zero terms because of the $\epsilon_{\mu\nu\lambda}$ tensor, so
\begin{eqnarray}
\mathcal{N}(G_{\sigma}) 
&=& -\frac{2 \cdot 6}{24\pi^2} \int d^3k  \frac{1}{((i\omega)^2-|\vec{h}_{\sigma}|^2)^2}
(\vec{h}_{\sigma} \cdot \frac{\partial \vec{h}_{\sigma}}{\partial k_{x}} 
 \times \frac{\partial \vec{h}_{\sigma}}{\partial k_{y}}) \nonumber \\
&=&
-\int \frac{d^2k}{4\pi} \frac{1}{|\vec{h}_{\sigma}|^3}
(\vec{h}_{\sigma} \cdot \frac{\partial \vec{h}_{\sigma}}{\partial k_{x}} 
 \times \frac{\partial \vec{h}_{\sigma}}{\partial k_{y}}) \nonumber \\
&=& -\int \frac{d^2k}{4\pi} 
\hat{h}_{\sigma} \cdot
\frac{\partial\hat{h}_{\sigma}}{\partial k_{x}} \times 
\frac{\partial\hat{h}_{\sigma}}{\partial k_{y}},
\end{eqnarray}
where the energy integral was done by computing the residue of the second order pole.

\section{\label{Sec:flux}Spin gauge flux $F^{s}_{\mu\nu}$ in terms of $\hat{N}$}
In the main text, we obtain the spin gauge field to be 
\begin{equation}
f_{\mu} = \frac{\sigma^{z}}{2} A^{s}_{\mu},
\end{equation}
where $f_{\mu} = -i U^{\dagger} \partial_{\mu} U$.

Therefore, we can write the spin gauge field in terms of the unitary matrix,
\begin{eqnarray}
A^{s}_{\mu} &=& \textrm{Tr} \left[ \sigma^{z} \cdot \frac{\sigma^{z}}{2} A^{s}_{\mu} \right] \nonumber \\
&=& \textrm{Tr} \left[ \sigma^{z} f_{\mu} \right] \nonumber \\
&=& -i\textrm{Tr} \left[ \sigma^{z} U^{\dagger} \partial_{\mu} U \right],
\end{eqnarray}
and we have
\begin{eqnarray}
F^{s}_{\mu\nu } &=& \partial_{\mu} A^{s}_{\nu} - \partial_{\nu} A^{s}_{\mu} \nonumber \\ 
&=& -i\textrm{Tr} \left[ \sigma^{z} (\partial_{\mu} U^{\dagger}) (\partial_{\nu} U) - \sigma^{z} (\partial_{\nu} U^{\dagger}) (\partial_{\mu} U)\right].
\end{eqnarray}
\begin{widetext}
Assume that the spin texture has the general form:
\begin{equation}
\hat{N}(\vec{x},t)=\left[ \sin \theta(\vec{x},t) \cos \phi(\vec{x},t), \sin \theta(\vec{x},t) \sin \phi(\vec{x},t), \cos \theta(\vec{x},t) \right],
\end{equation}
where $\theta(\vec{x},t)$ and $\phi(\vec{x},t)$ can be any function of position and time.
Then, we have the unitary matrix
\begin{equation}
U(\vec{x},t)=\left(
\begin{array}{cc}
\cos \frac{\theta(\vec{x},t)}{2} & -\sin \frac{\theta(\vec{x},t)}{2} e^{-i\phi(\vec{x},t)} \\
\sin \frac{\theta(\vec{x},t)}{2} e^{i\phi(\vec{x},t)} & \cos \frac{\theta(\vec{x},t)}{2}
\end{array}
\right),
\end{equation}
\begin{equation}
\partial_{\mu}U^{\dagger}(\vec{x},t)=\left(
\begin{array}{cc}
-\frac{1}{2} \sin \frac{\theta}{2} \partial_{\mu} \theta &
e^{-i\phi} ( \frac{1}{2} \cos \frac{\theta}{2} \partial_{\mu} \theta 
- i \sin \frac{\theta}{2} \partial_{\mu} \phi  ) \\
e^{i\phi} ( -\frac{1}{2} \cos \frac{\theta}{2} \partial_{\mu} \theta 
- i \sin \frac{\theta}{2} \partial_{\mu} \phi  ) 
 & -\frac{1}{2} \sin \frac{\theta}{2} \partial_{\mu} \theta
\end{array}
\right),
\end{equation}
and
\begin{equation}
\partial_{\nu}U(\vec{x},t)=\left(
\begin{array}{cc}
-\frac{1}{2} \sin \frac{\theta}{2} \partial_{\nu} \theta &
e^{-i\phi} ( -\frac{1}{2} \cos \frac{\theta}{2} \partial_{\nu} \theta 
+ i \sin \frac{\theta}{2} \partial_{\nu} \phi  ) \\
e^{i\phi} ( \frac{1}{2} \cos \frac{\theta}{2} \partial_{\nu} \theta 
+ i \sin \frac{\theta}{2} \partial_{\nu} \phi  ) 
 & -\frac{1}{2} \sin \frac{\theta}{2} \partial_{\nu} \theta
\end{array}
\right),
\end{equation}

where we have suppressed the arguments of $\theta(\vec{x},t)$ and $\phi(\vec{x},t)$.
 
Therefore, we can calculate the product of the last two matrices, and express the spin gauge flux as
\begin{eqnarray}
F^{s}_{\mu\nu} 
&=& -i \left[\frac{i}{2} \sin \theta (\partial_{\mu} \theta \partial_{\nu} \phi-\partial_{\nu} \theta \partial_{\mu} \phi) \right] \times 2 \nonumber \\
&=& \sin \theta \left( \partial_{\mu} \theta \partial_{\nu} \phi - \partial_{\nu} \theta \partial_{\mu} \phi \right).
\end{eqnarray}

In addition, we can also write $\hat{N} \cdot (\partial_{\mu}\hat{N} \times \partial_{\nu}\hat{N})$ in terms of $\theta(\vec{x},t)$ and $\phi(\vec{x},t)$,
\begin{eqnarray}
&&\hat{N} \cdot (\partial_{\mu}\hat{N} \times \partial_{\nu}\hat{N}) \nonumber \\
&=& \left|
\begin{array}{ccc}
\vspace{0.1in}
\sin \theta(\vec{x},t) \cos \phi(\vec{x},t) & \sin \theta(\vec{x},t) \sin \phi(\vec{x},t) & \cos \theta(\vec{x},t) \\
\left[ \cos \theta(\vec{x},t) \cos \phi(\vec{x},t) \partial_{\mu}\theta(\vec{x},t) \right. 
 &
\left[ \cos \theta(\vec{x},t) \sin \phi(\vec{x},t) \partial_{\mu}\theta(\vec{x},t) \right.
 & -\sin \theta (\vec{x},t) \partial_{\mu}\theta(\vec{x},t) \\
\left.- \sin \theta(\vec{x},t) \sin \phi(\vec{x},t) \partial_{\mu}\phi(\vec{x},t)\right]
 &
\left.+ \sin \theta(\vec{x},t) \cos \phi(\vec{x},t) \partial_{\mu}\phi(\vec{x},t)\right]
 &  \vspace{0.1in} \\
\left[ \cos \theta(\vec{x},t) \cos \phi(\vec{x},t) \partial_{\nu}\theta(\vec{x},t) \right. 
 &
\left[ \cos \theta(\vec{x},t) \sin \phi(\vec{x},t) \partial_{\nu}\theta(\vec{x},t) \right.
 & -\sin \theta (\vec{x},t) \partial_{\nu}\theta(\vec{x},t) \\
\left.- \sin \theta(\vec{x},t) \sin \phi(\vec{x},t) \partial_{\nu}\phi(\vec{x},t)\right]
 &
\left.+ \sin \theta(\vec{x},t) \cos \phi(\vec{x},t) \partial_{\nu}\phi(\vec{x},t)\right]
 &  \vspace{0.1in}
\end{array}
\right|, \nonumber \\
&=&
\sin \theta \left( \partial_{\mu} \theta \partial_{\nu} \phi - \partial_{\nu} \theta \partial_{\mu} \phi \right),
\end{eqnarray}
where, again, we suppressed the arguments of $\theta(\vec{x},t)$ and $\phi(\vec{x},t)$. Finally, we obtain
\begin{equation}
F^{s}_{\mu\nu } = \hat{N} \cdot (\partial_{\mu}\hat{N} \times \partial_{\nu}\hat{N}).
\end{equation}
\end{widetext}

%

\end{document}